\documentclass[12pt,preprint]{aastex}

\def\gs{\mathrel{\raise0.35ex\hbox{$\scriptstyle >$}\kern-0.6em
\lower0.40ex\hbox{{$\scriptstyle \sim$}}}}
\def\ls{\mathrel{\raise0.35ex\hbox{$\scriptstyle <$}\kern-0.6em
\lower0.40ex\hbox{{$\scriptstyle \sim$}}}}

\shorttitle{Molecular  gas  in  HR\,10}
\shortauthors{Papadopoulos \& Ivison}

\begin{document}

\title{Low-excitation gas in HR\,10: possible 
      implications for estimates of metal-rich H$_2$ mass at high
      redshifts}

\author{P.\ P.\ Papadopoulos}
\affil{Astrophysics Division, Space Science Department of ESA,
             ESTEC, Postbus 299, NL-2200 AG, Noordwijk,
             The Netherlands; ppapadop@rssd.esa.int}

\and

\author{R.\ J.\ Ivison}
\affil{Astronomy Technology Centre, Royal Observatory, Blackford Hill,
Edinburgh EH9 3HJ, UK; rji@roe.ac.uk}

\begin{abstract}
We examine the physical conditions of the molecular gas in the
extremely red object (ERO), HR\,10 (J164502+4626.4, or [HR94]\,10) at
$z=1.44$, as constrained by the recent detection of CO $J+1
\rightarrow J, J=1, 4$ lines.  The line ratios of such widely spaced
CO transitions are very sensitive probes of gas excitation; in the
case of HR\,10, they provide a rare opportunity to study the state of
H$_2$ gas in a gas-rich starburst at a cosmologically significant
distance.  Contrary to earlier claims, we find the CO $J=5-4$
transition to be of low excitation with a brightness temperature ratio
of ($5-4$)/($2-1$) $\sim$ 0.16.  Depending upon how often such
conditions prevail in intense starbursts, this may have serious
consequences for molecular gas mass estimates and the detectability of
high-$J$ CO lines from similar objects at high redshifts ($z\ga 3$).
\end{abstract}


\section{Introduction}

A defining characteristic of highly obscured active galactic nuclei
and star-forming galaxies is their red optical/infrared (IR) colours.
The first few EROs (usually defined such that $R-K>6$ or $I-K>5$) were
found soon after the advent of near-IR array detectors (Elston, Rieke
\& Rieke 1988) but were viewed as little more than curiosities.  It
has become  clear,  however, that  while   EROs provide  a  negligible
contribution  to the  optical extragalactic background,  they may host
power  sources which dominate the  background at other wavelengths, in
particular in the submillimetre (submm) and hard X-ray wavebands.

Recent  work suggests that  EROs  form a  bi-modal population: a large
fraction, possibly around  half at  $K>20$,   can be identified   with
obscured starbursts (e.g., Dey  et al.\ 1999); the remainder represent
the end products of these starbursts --- massive ellipticals at $z
\sim$ 1--2 (e.g., Dunlop et al.\ 1996) whose colours are due to their
evolved stellar populations.

Submm surveys with SCUBA (Holland et al.\ 1999) have demonstrated that
a large fraction of the extragalactic submm background originates in
ultra-luminous IR galaxies (ULIRGs, $L_{\rm FIR}\ga
10^{12}$\,L$_{\odot}$) at high redshifts (Smail, Ivison \& Blain
1997), a population which provides well below 1 per cent of the local
luminosity density and which must have evolved dramatically, having
provided closer to half of the total background at earlier times. Like
the local examples of ULIRGs, many submm sources are distant, gas-rich
mergers, the probable progenitors of luminous present-day ellipticals
(Ivison et al.\ 1998, 2000; Smail et al.\ 1998, 2000; Frayer et al.\
1998, 1999; Eales et al.\ 1999).  Unsurprisingly, given the bi-modal
population discussed earlier, several have been identified with EROs
(Smail et al.\ 1999). Submm data, in fact, have proved to be an ideal
discriminant between the obscured starburst and old elliptical EROs
(Dey et al.\ 1999).

HR\,10, an optical/IR-selected galaxy at $z=1.44$ (Hu \& Ridgway 1994;
Graham \& Dey 1996), was the first ERO to be detected at submm
wavelengths (Cimatti et al.\ 1998; Dey et al.\ 1999) and to be thus
revealed unequivocally as an extremely luminous, dust-obscured
starburst ($\rm L_{\rm FIR}\sim 10^{12}\ L_{\odot }$), a fact that
would have been difficult to surmise from its non-descript
optical/near-IR appearance (Dey et al.\ 1999).

The  proper estimation  of  molecular  gas   mass in such    gas-rich,
IR-luminous objects is of central importance since the gas fuels their
often   enormous  ($\sim 10^3$ M$_{\odot  }$~yr$^{-1}$) star-formation
rates (e.g., Solomon et al.\ 1992; Dunlop et al.\ 1994; Rowan-Robinson
2000 and references therein).  Comparing $M$(H$_2$) with the dynamical
mass allows the determination of  the evolutionary status of a galaxy,
while a comparison  of its dynamical  mass with that  of a present-day
spiral or elliptical can point towards its  possible descendent at the
current   epoch.   Finally,   since  a  dust-mass   estimate is  often
available, the resulting  gas-to-dust ratio is a significant indicator
of the   metal enrichment and  chemical evolution   of galaxies in the
early Universe (e.g., Frayer \& Brown 1997).

In this paper we examine the  physical conditions of the molecular gas
in  the ERO,  HR\,10  at  $z=1.44$, and  their  effect on  H$_2$  mass
estimates.  Finally, to the extent that such conditions are frequently
encountered in  starburst  environments,  there   may be  far-reaching
consequences  for the detection and interpretation  of high-J CO lines
emitted from  starbursts at high redshifts  --- for the prioritisation
of  observing bands and  receivers  for the  Atacama  Large Millimetre
Array (ALMA),  for example. Throughout  this work, we  adopt $q_{\circ
}=0.5$ and $H_{\circ }=75\rm \ km\ s^{-1}\ Mpc^{-1}$.

\section{Physical properties of H$_2$ in HR\,10}

The detection  of the  two  widely  spaced CO transitions  in   HR\,10
($J=2-1, J=5-4$, Andreani et al.\ 2000) offers a unique opportunity to
examine  the  physical  conditions  of the   molecular  gas using  the
excitation-sensitive ratio  of their brightness temperatures.  This is
a   rare occurrence  for high-redshift objects   ($z\ga   1$) since it
requires a fortuitous combination of redshift and available receivers.
For low-redshift objects ($z\la 0.1$), systematic studies of this sort
are hindered    by  large atmospheric  attenuation   and  low  antenna
efficiencies for frequencies $\nu > 400$\,GHz, where the CO $J+1
\rightarrow J, J+1 > 3$ transitions lie and, until the commissioning
of a new generation of wide-band heterodyne receivers, by the
bandwidth and sensitivity limitations of available instrumentation.

The detection of the  CO $J=2-1$ transition is  particularly important
since, together with CO $J=1-0$, is  the most widely studied molecular
line in the local Universe.  The critical  densities of CO $J=1-0$ and
$J=2-1$ (i.e.  the densities at which the collisional  rate out of the
upper level and its  radiative de-excitation become equal) are $n_{\rm
crit}\sim 2.4\times  10^2-2.4\times 10^3$\,cm$^{-3}$ (for  $\rm T_{\rm
k}=100$ K;  see e.g. Kamp \& van  Zadelhoff 2001), and $\Delta E/k\sim
5-10$\,{\sc  k}.  These  minimal  excitation requirements allows these
transitions to trace the bulk of the  metal-enriched molecular gas and
there  are  several surveys measuring their  relative  strengths for a
wide variety of environments (e.g.,  Braine  \& Combes 1992;  Hasegawa
1997; Papadopoulos \& Seaquist 1998).

The CO $J+1 \rightarrow J$ line luminosity in units of {\sc
k}\,km\,s$^{-1}$\,pc$^{2}$ is given by

\noindent
\begin{equation} 
L_{\rm CO}(J+1\rightarrow J) = \frac{c^2}{2\ k\ \nu_{J+1,J}^2}
\left[\frac{D^2_{\rm L}}{\ 1+z\ }\right] \int _{\Delta v} S_{\nu _{\rm
obs}} d v,
\end{equation}

\noindent
where $D_{\rm L}=2\ c\ H_{\circ}^{-1} (1+z -\sqrt{1+z})$ is the
luminosity distance, and $\nu_{J+1, J}$, $S_{\nu_{\rm obs}}$, are the
rest-frame frequency and the observed flux density of the CO $J+1
\rightarrow J$ transition (see, e.g., Ivison et al.\ 1996).

Andreani et al.\ (2000) have reported almost equal velocity-integrated
flux  densities for  the  $J=5-4$ and  $J=2-1$ transitions, hence  the
area/velocity-integrated brightness temperature ratio will simply be

\noindent
\begin{equation}
\frac{L_{\rm CO}(5-4)}{L_{\rm CO}(2-1)}\sim \left( \frac{\nu
_{21}}{\nu _{54}}\right)^2=\left( \frac{2}{5}\right)^2= 0.16
\end{equation}

\noindent
which is $\sim 4$ times lower than  their reported value and indicates
the low excitation of the CO $J=5-4$ line.  The new value is bracketed
by the global value of $\sim 0.07$ for  the Milky Way, as derived from
{\em Cosmic Background  Explorer} measurements (Wright et al.\  1991),
and  $\sim  0.35-1$ expected  for the  conditions  in  the inner $\sim
$500\,pc of  the archetypal starburst  M\,82 (using data from Seaquist
\& Frayer 2000; Mao et al.\ 2000 and references therein).
 
The  surprising fact  is not that   the aforementioned  line  ratio is
higher for the starburst environment of  HR\,10 than for the quiescent
one of the Milky  Way but that it  is at least  a  factor of $\sim  2$
lower  than in  the prototype  starburst galaxy,  M\,82.  The  case of
HR\,10 shows  that  very different  excitation  conditions may  indeed
prevail in  such   systems.    It is worth    noting  that   such  low
(5--4)/(2--1) ratios are  associated with the low-excitation gas phase
present in  M\,82  (e.g., Mao  et al.\  2000  and references therein),
suggesting that this phase is dominant in HR\,10.

The estimate of H$_2$ mass  in the Galaxy  and other galaxies utilises
the bright  CO $J=1-0$ emission  and the so-called Galactic conversion
factor,  $X_{\rm  CO}$ (e.g., Young  \&  Scoville  1982, 1991; Bloemen
1985; Dickman, Snell \& Schloerb 1986; Bryant \&  Scoville 1996).  For
an observed CO $J+1 \rightarrow J$ line, the  H$_2$ mass is then given
by

\noindent
\begin{equation}
M({\rm H}_2)=X_{\rm CO}\ R_{J+1, J}^{-1} \ L_{\rm CO}(J+1, J)
\end{equation}

\noindent
where $R_{J+1, J}$ is ratio of the CO $J+1 \rightarrow J$ and $J=1-0$
line luminosities.

The  range  of  $R_{21}$  observed   in  a variety    of  Galactic and
extragalactic environments  is $\sim  0.5-1$ (e.g., Braine   \& Combes
1992; Hasegawa 1997; Papadopoulos \& Seaquist 1998) hence the expected
range of  $R_{54}$ for  HR\,10  is $\sim 0.08-0.16$  (see Equation~2),
significantly lower than $\sim 1$,   the value for an optically  thick
and thermalized line.

Unlike strongly lensed objects, where   differential lensing may  bias
the dust continuum and line ratios towards a warmer dust/gas component
(Blain 1999; Papadopoulos  et   al.\    2000), the  lack  of    strong
gravitational lensing  in   HR\,10 (Dey  et al.\ 1999)   allows a more
straightforward interpretation  of  the observed line   ratio.  In the
absence  of more data,  we  examine the physical conditions compatible
with the observed (5--4)/(2--1) line ratio assuming a single gas phase
and  using  a  large-velocity gradient   (LVG)  code (e.g., Richardson
1985).  The code was run for the typical range of temperatures $T_{\rm
kin}=30-60$\,{\sc k} that characterise  the warm-dust component (where
the  bright   CO    emission  also arises)   in    IR-bright  galaxies
(e.g.  Devereux  \& Young 1990   and references  therein).   A typical
solution    is      $T_{\rm     kin}=45$\,{\sc k}    and       $n({\rm
H}_2)=10^3$\,cm$^{-3}$  ($T_{\rm dust}\sim 40$\,{\sc  k}, Dey  et al.\
1999).      Such a  diffuse,    low-excitation gas    phase with  such
temperatures is  not unusual in  extreme starburst systems  (Downes \&
Solomon 1998); it is present (but not dominant)  in much less luminous
starbursts like M\,82 (Hughes, Gear, \& Robson 1994; Mao et al.\ 2000)
and is   probably the product  of their  UV-intense  and kinematically
violent environments (Aalto et al.\ 1995).

For these conditions $R_{54}\sim 0.14$,  while higher transitions have
$R_{65}\sim   0.035$  and $R_{\rm J+1,   J}   \la 5\times 10^{-3}$ for
$J+1>6$.  The low excitation of the $J=5-4$  line in particular is due
to the low density rather than  temperature since, for optically thick
and thermalized gas, such a low ratio implies $T_{\rm kin} < 10$\,{\sc
k},  much  lower than  the observed  dust temperature.    The claim by
Andreani et al.\ that $T_{\rm kin}$ is usually significantly less than
$T_{\rm dust}$ is not valid; usually they are  similar, except for the
photodissociation  regions  (PDRs) in  the surfaces  of UV-illuminated
molecular clouds where $T_{\rm kin} > T_{\rm dust}$ (e.g., Hollenbach
\& Tielens 1999). Here we must stress that all other LVG solutions
compatible with the low (5--4)/(2--1) ratio result to $\rm R_{\rm J+1,
J} \la 0.05$ for $\rm J+1\geq 6$.

\section{Consequences for high-redshift galaxies}

At high  redshifts  ($z\ga 3$), the  usual  instruments of  choice for
detecting CO lines are millimetre-wave interferometers. These are only
capable  of  accessing high-$J$  (i.e.\  $J>2$) CO $J+1 \rightarrow J$
lines.  In  such cases, the  obvious consequence of the low excitation
of such lines  is that they become  difficult to detect, even if there
is a  large quantity of molecular  gas.  Equivalently, any  H$_2$ mass
estimates  based on the assumption  of optically thick and thermalized
high-$J$ CO lines  (e.g., Evans et al.\  1996; van Ojik et al.\  1997)
may underestimate the mass  present, or  its  upper limit, by  a large
factor ($\ga 10$).

In the case of HR\,10, scaling its CO $J=5-4$ flux to $z=4$
(reasonably typical for the large samples of quasars, radio galaxies,
Lyman-break and submm-selected galaxies that are now being assembled
-- e.g., Steidel et al.\ 1999; Smail et al.\ 2000) yields a
velocity-integrated flux density of $\sim 0.8$\,Jy\,km\,s$^{-1}$ (from
Equation~1), received at $\nu_{\rm obs} \sim 115$\,GHz.  This is a
rather weak signal. While still detectable in a dedicated observing
run of a few tens of hours with a sensitive present-day
millimetre-wave interferometer, it would evade detection in a typical
survey for such lines in full samples of high-redshift galaxies (e.g.,
van Ojik et al.\ 1997). Moreover, an estimate of $M$(H$_2$) based on
the CO $J=5-4$ line only, and the assumption that it is optically
thick and thermalized, would yield a value $\sim 6-12$ times too low
(Equation~3, $R_{54}\sim 0.08-0.16$).  The previous simple excitation
analysis suggests that CO $J+1 \rightarrow J$ ( with $J+1 \geq 5$)
lines can be poor global mass tracers in such environments,
underestimating the H$_2$ mass by factors of $\ga 10$, depending on
the particular line used and the prevailing physical conditions.

For the least excitation-biased mass estimates of metal-enriched
molecular gas it is clearly important to observe the lowest two
transitions of CO.  At high redshifts this is currently possible only
in cases where there is a fortuitous combination of redshift and
available receivers.  Sadly, this precludes a systematic inventory of
molecular gas mass at $z\ga 3$ with present instruments, though the
Green Bank Telescope, the upgraded Very Large Array (VLA) and ALMA
will provide the necessary instantaneous bandwidth (velocity
coverage), tuning range (redshift coverage) and sensitivity to detect
many CO lines at such redshifts.  Fig.~1 demonstrates that several CO
line ratios will be available for galaxies at virtually all redshifts,
with a small gap at $z\sim 2.7-3.4$ ($z\sim 2.7-4.1$ if one considers
only ALMA).  For the CO(1--0) and CO(2--1) lines, however, ALMA
provides the only coverage at $z\sim 2.0-2.7$, a potentially important
range for a number of known galaxy populations (Steidel et al.\ 1999;
Ivison et al.\ 1998, 2000; Smail et al.\ 2000), and to do this ALMA
requires the so-called `Band~1' receivers (31.3--45\,GHz).

Currently, the commissioning of a new generation of sensitive
receivers with wider tuning ranges on mm/sub-mm telescopes around the
world will enable the extension of multi-line studies of nearby
starbursts ($\rm L_{\rm FIR}\sim 10^{9}-10^{10} L_{\odot }$) towards
more luminous systems ($\rm L_{\rm FIR}\sim 10^{11}-10^{12} L_{\odot
}$) at larger distances.  This will reveal whether a low-excitation
gas phase dominating the emission of the $ ^{12}$CO rotational lines
is indeed more frequently encountered at higher FIR luminosities and
will yield a firm assessment of the expected luminosities of
higher-$J$ lines in luminous starbursts found at high redshift.

\section{Conclusions}

We  have  presented a  brief  analysis  of the  physical conditions of
molecular gas in the $z=1.44$  extremely red object, HR\,10, using the
CO $J=2-1,  J=5-4$   lines.  HR\,10 provides   a  rare opportunity  to
investigate   low-$J$ {\it  and} high-$J$   CO  lines  in an   intense
starburst at a cosmologically significant distance.  Its case can thus
be instructive  for similar  environments in  objects at still  higher
redshifts.  Our conclusions are as follows:

\begin{itemize}
\item
Contrary to earlier claims, we find the CO $J=5-4$ line to be of low
excitation, with a brightness temperature ratio of $(5-4)/(2-1) \sim
0.16$, likely due to low densities rather than low temperatures.
\item
The CO $J+1\rightarrow J, J+1 \geq 5$ emission lines from similar
galaxies at $z\ga 3$ would be hard to detect using current
millimeter-wave arrays.  If detected, the usual assumption made for
their excitation, namely that they are thermalized and optically
thick, may underestimate H$_2$ gas masses by a factor $\ga 10$.
\end{itemize}

This analysis may have some bearing  on the relative priority given to
the so-called `Band~1' frequencies/receivers (31.3--45\,GHz) for ALMA.
It is at these frequencies that one would detect the lowest-excitation
CO $J=1-0$     transition   from  a substantial     fraction    of the
submm-selected  galaxy population   (`SCUBA galaxies'),  systems which
appear to have a median redshift of 2.5--3  (Smail et al.\ 2000).  The
upgraded VLA may also have a significant role to play;  it will have a
similar {\em effective} aperture to ALMA at those frequencies, but the
expected phase stability at Atacama  and the $2\times$ larger field of
view means that ALMA is likely to be the interferometer of choice.

\acknowledgments

We thank the referee Christian Henkel for useful comments that
improved the present work and acknowledge support by a Joint Project
grant from the Royal Society.

\clearpage

%
%
\begin{figure}
\plotone{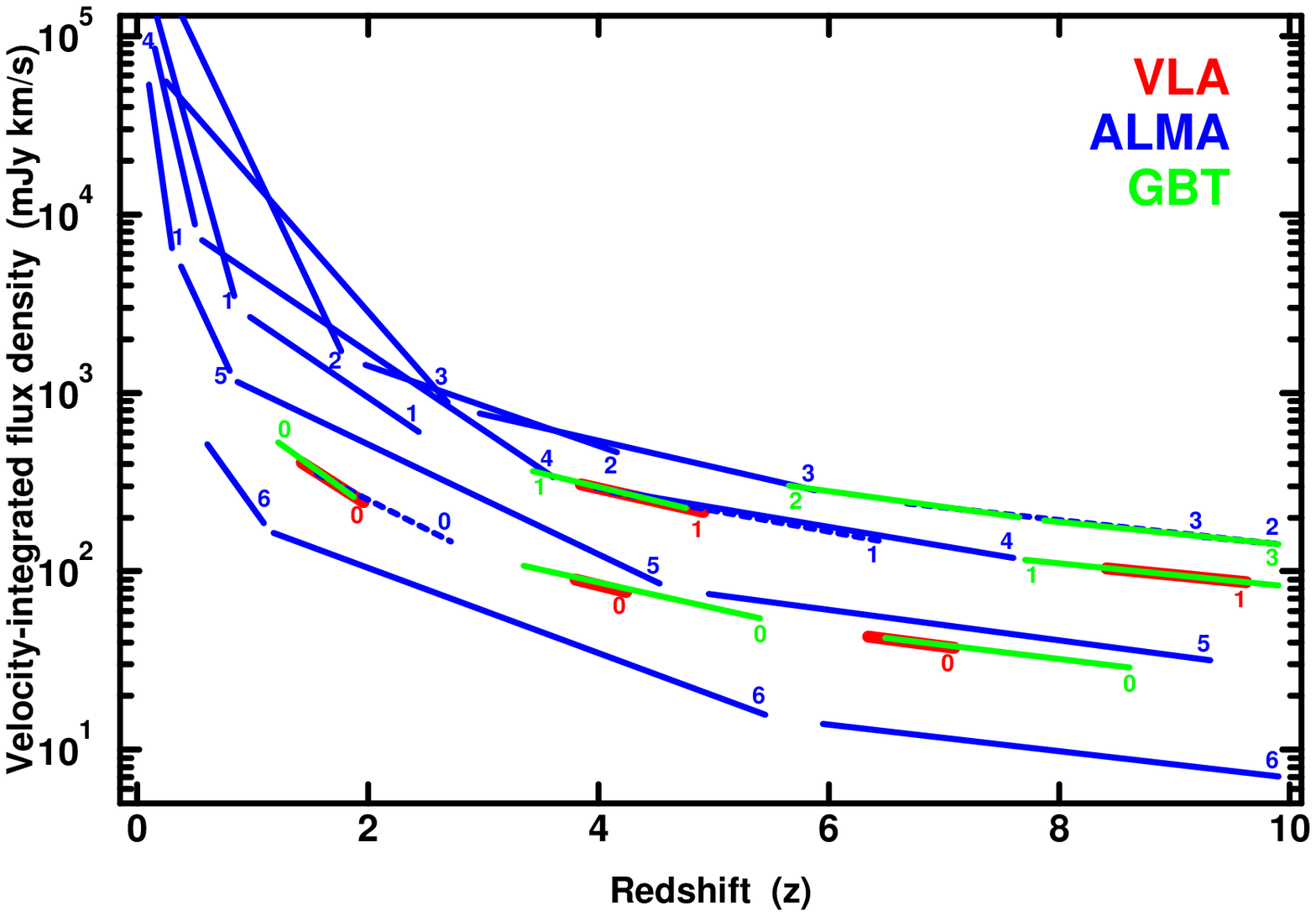}
\caption{The expected velocity-integrated flux density, $\int S_{\nu
_{\rm obs}} (\rm J+1, J)\ dV$, ($J$ is labeled, $J=0-6$) emitted by
$10^{11}$\,M$_{\odot}$ of metal-enriched H$_2$ gas ($\rm Z/Z_{\odot
}=1$) with properties identical to those of HR\,10.  Each line
represents the velocity-integrated flux density appropriate for the
redshifts where each $J+1 \rightarrow J$ transition is available to
the receivers at a particular telescope.  For the VLA, the bands are
taken to be the regions where current receivers have at least 50 per
cent of their optimum sensitivity (Perley 2000).  Plans for the
Extended VLA include continuous frequency coverage from 1.1 to
50\,GHz, which will provide coverage of low-$J$ CO transitions at
$z\ge 1.3$.  For ALMA, the definitions of `Bands 1--10' were taken
from the ALMA Construction Project Book
(www.tuc.nrao.edu/$\sim$demerson/almapbk/construc/cons\_toc.htm), and
Band 1 coverage is shown by dashed lines; for the GBT, the bands were
taken from the Short Guide to the Green Bank Telescope
(www.nrao.edu/GBT/proposals/short\_guide.html).}
\end{figure}

\end{document}